# "AI+建筑学"模块化融合设计课教学探索：以浙江大学建筑设计Ⅲ/Ⅳ课程为例

王嘉琪 WANG Jiaqi 1,2

蓝逸 LAN Yi 1

陈翔 CHEN Xiang 1,2,3 （通讯作者）

1 浙江大学建筑工程学院建筑学系（杭州，310058）

2 浙江大学平衡建筑研究中心（杭州，310058）

3 浙江大学建筑设计研究院有限公司（杭州，310028）

## 摘要

本文以浙江大学 2024-25 学年本科三年级建筑设计核心课程教学实验，探索 AI 技术融入本科建筑教育的路径。课程采用"20 课时 AI 技能模块+课堂 AI 伦理研讨"的双模块框架，在不改变原有教学结构的前提下，引入深度学习模型、大语言模型、人工智能生成图像内容、LoRA 微调及 ComfyUI 等技术，并配备专职技术导师。教学实验证实，分阶段指导、技术伦理平衡、平台保障等制度支持是关键。这种融合模式提升了学生的数字适应力和策略思维，同时引发对 AI 协作中的著作权、算法偏见等问题的思考，为平衡技术训练与批判意识的设计教学提供了可复制的参考方案。

## 关键词

多模态人工智能；人工智能伦理；建筑学；模块化；本科教学改革；创新复合型人才

Exploring the Modular Integration of "AI + Architecture" Pedagogy in Undergraduate Design Education: A Case Study of Architectural Design III/IV Courses at Zhejiang University

ABSTRACT

This study investigates AI integration in architectural education through a teaching experiment in Zhejiang University's 2024-25 grade three undergraduate design studio. Adopting a dual-module framework (20-hour AI training + embedded ethics discussions), the course introduced deep learning models, LLMs, AIGC, LoRA, and ComfyUI while maintaining the original curriculum structure, supported by dedicated technical instructors. Findings demonstrate the effectiveness of phased guidance, balanced technical-ethical approaches, and institutional support. The model improved students' digital skills and strategic cognition while addressing AI ethics, providing a replicable approach combining technical and critical learning in design education.

自 20 世纪初以来，建筑设计技术经历了从手工绘图的感性训练，到计算机制图与参数化建模工具的广泛应用，再到人工智能（AI）逐步介入、协同与增强设计过程的演进转型，展现出日益加速的技术迭代趋势。围绕建筑设计构建专属的 AI 协同设计路径、知识图谱数据库与增强设计平台，正日益成为提升行业核心竞争力的关键方向[1]。在此背景下，建筑设计企业已率先将 AI 工具应用于实际项目。福斯特合伙人事务所（Foster + Partners）于 2019‑2022 年间开发并部署智能设计工具 Hydra，生成方案逾 24 万、完成模拟达 130 万次，显著提升前期设计效率[2]。扎哈·哈迪德建筑事务所（Zaha Hadid Architects）自 2022 年起则引入 Midjourney、Gendo 及自研工具，将竞赛阶段设计效率提升 2 至 3 倍，渲染时间缩短至 20%，显著增强了其竞赛优势[3]。

AI 技术正重塑建筑设计教育的教学模式、表达方式与认知体系，传统以手工操作与绘图技能为核心的教学方式，正逐步被强调建模逻辑、生成机制与反馈迭代的新范式所替代[4,5]。但有研究指出，高校在 AI、参数化设计、BIM（建筑信息模型）等工具的系统训练方面仍显薄弱，教学内容与行业实践之间存在一定脱节[6]。然而，在变革加速的当下，教学改革尤需审慎推进，警惕技术逻辑对设计思维的误导。因此，AI 融入建筑教育宜以教学实验先行，通过实践不断验证与调整改革路径。

目前，国内部分高校已在建筑学专业启动 AI 教学实验。同济大学于 2023 年开设《AI 辅助建筑编程与设计》研究生课程，系统引入 Midjourney、ChatGPT、Stable Diffusion 等主流人工智能生成内容（AIGC）工具，24 名学生参与后普遍给予积极评价，认为其"创新能力"与"工作效率"显著增强，但也暴露出设计表达不可控、应用碎片化等问题[7]。清华大学于同年针对建筑学本科三年级开设《AI 生成式影像》跨学科课程实验，引导学生利用 Midjourney、Runway 等 AIGC 工具全流程创作影像短片，探索交叉学科课程教学路径[8]。东南大学于 2024 年针对建筑学本科四年级开设《可变住宅》设计课，综合使用各类生成式算法及 Processing（基于 Java 的可视化编程环境）、MOEA（多目标进化算法）等多种 AI 工具，实现住宅构造与户型组合的自

动生成[9]。东北大学同年将 Stable Diffusion 引入建筑学二年级设计课，虽提升了方案迭代效率，但也面临生成结果不可控、学生 AI 掌控力不足等问题[10]。

放眼国际，国外高校亦积极回应技术浪潮，系统构建融合 AI 技术的课程体系与教学平台。哈佛大学设计研究生院（GSD）已将 AI 技术纳入常规设计支持手段，信息技术部门设立专属平台，持续更新 AI 工具指南、教学政策与伦理规范，形成较为完善的 AI 教学支持体系[11]。2025 年初，GSD 还开设了面向非技术背景群体的短期课程《AI, Machine Learning, and the Built Environment（人工智能、机器学习与建成环境）》，重点介绍 AI 与机器学习在不动产、建筑、景观与城市设计等领域的应用[12]．哥伦比亚大学建筑、规划与保护研究生院（GSAPP）自 2022 年起陆续设立了《Spatial AI（空间人工智能）》《AI for Existing Buildings（既有建筑的人工智能应用）》等选修课程，聚焦 AI 与建筑设计的融合[13,14]。代尔夫特理工大学（TU Delft）亦推出线上慕课《AI in Architectural Design: Introduction（建筑设计中的人工智能导论）》，向全球开放[15]。

综上所述，当前国内外高校均在积极探索以 AIGC 为代表的 AI 技术在建筑教育中的应用。国外主要聚焦研究生阶段，课程多为讲授型，面向本科设计课的代表性成果尚属稀缺；国内虽有面向本科生的教学实践，但其教学内容与既有体系差异显著，尚处于调整磨合期。可见，AI 正在重塑建筑教育的理念与方法，而课程体系与教学机制的重构仍处起步阶段，亟须更多具体实践予以验证与完善。为此，本文以浙江大学建筑学系 2024-2025 学年《建筑设计Ⅲ/Ⅳ》课程中开展的"AI+建筑学"教学实验为例，尝试探讨 AI 知识模块化融入核心设计课的路径与成效，以期为 AI 技术常态化融入建筑设计课教学提供案例参考。

## 1.课程简介及其组织模式

### 1.1 课程改革背景

近十年来，浙江大学建筑学系持续推动新兴技术与建筑教育的深度融合。2018 年人工智能与机器人建造实验室的建成，为开展计算设计与 AI 辅助设计领域的教学科研工作提供了重要平台[16]。在此背景下，建筑学系对 2024 级新生培养方案进行了全面修订，不仅在必修课程体系中新增《人工智能基础（B）》课程，更通过增设选修课程、重构课程模块以及开展短期强化工作坊等多元化形式，系统性地推进人工智能与数字化设计技术的教学改革。

在既往教学改革中，本系主要借鉴国内主流模式，通过增设技术类选修课程逐步引入新兴技术教学内容。2024 年实施的新一轮改革着力构建"基础-进阶-创新"三级培养体系，在本科三个年级同步推进智能技术与核心设计课程的融合实验：一年级重点培养数字化基础能力，涵盖 AutoCAD 工程制图、SketchUp/Rhino 三维建模及 D5/Enscape 渲染等模块；二年级着重强化计算性设计思维，整合 Grasshopper 参数化生成、Climate Consultant 气候分析及 Depthmap 空间句法解析等方法体系；三年级着力探索 AI 设计创新，引入深度学习模型、大语言模型及人工智能生成内容等技术集群。（图 1）

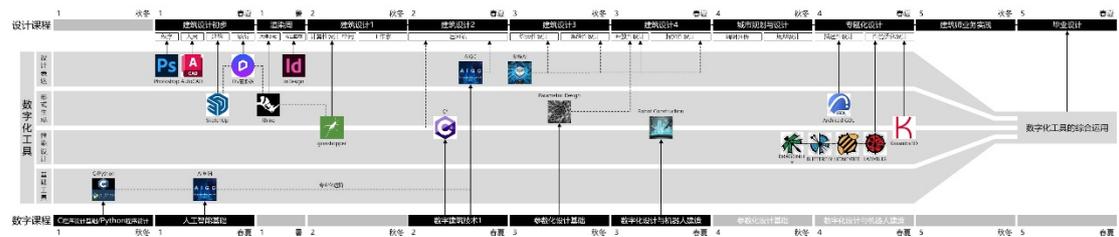

图 1 浙江大学建筑学本科五年数字化工具教学组织

### 1.2 建筑设计Ⅲ/Ⅳ课程简介

建筑设计系列课程贯穿浙江大学建筑学本科五年制教学，采用"3+1+1"的设计教学体系[17]。前三年强调基础专业素养，从一年级的基于空间和构成的初步训练，到二年级的针对建筑基本问题的切片式训练，到三年级的基于复杂建筑问题的综合训练；第四年是强调个性化设计教学的专题化设计训练；五年级则是基于独立综合设计能力培养的、强调与实战对接的毕业设计。

作为建筑学专业核心设计课程的阶段性总结，三年级《建筑设计Ⅲ/Ⅳ》课程重点培养学生应对复杂建筑设计问题的能力，以及综合运用各类工具技术的专业素养。该课程为期一学年，分为约束性设计（建筑改造方向）、系统性设计（复合约束方向）、开放性设计（概念探索方向）、探究性设计（城市更新方向）四个设计任务。

本次教学实验在完整保留传统课程框架的基础上，重点探索AI技术与设计教学的深度融合路径。实验主要针对三个现实问题展开：第一，建筑学教育中AI知识教学尚处探索阶段，缺乏成熟案例；第二，学生设计方法论仍囿于传统范式，亟需构建人机协同的设计思维，同时培养其对AI伦理价值的批判性认知能力；第三，学生对AI技术的理解较为碎片化，需要通过实践场景培养其系统性整合应用能力。

### 1.3 课程组织

针对上述这些问题，课程创新性地融入了模块化AI训练单元，设置AI技术与AI伦理模块（表1）。AI技术模块采用"20课时集中授课+云端实训"的混合教学模式，逐步引入AIGC、DCNN（深度卷积神经网络模型）、LLM（大语言模型）、LoRA（低秩自适应微调模型）与ComfyUI（节点式图像生成工作流）等核心技术。教学依托校级智云课堂平台与云计算基础设施，同时支持本地与云端部署，并持续更新开源模型与数据集资源库，学生可自主选择线上或线下学习。AI伦理模块则采用随堂讨论形式，不额外增加学时，重点探讨数据隐私与著作权、人机协同的权责边界、模型风格化创作的文化偏见及模型算法偏误四大议题。

在课程组织上，AI教学模块采用渐进式融入策略，根据学生掌握程度，结合四个设计任务的教学节奏，逐步嵌入设计任务的"调研-设计-表现"全流程：首先在设计阶段介入，后续延申至设计-表现阶段，并在最终拓展至全部三个阶段，通过反复强化训练实现从设计阶段单点应用到全流程整合的能力提

升。这种分阶段、多模块的"AI+建筑学"模式，既可确保传统设计教学的连贯性，又能达成 AI 知识与专业内容的深度协同创新。

表 1 AI 专题模块与原教案融合方案

| 周次 | 总学时 | 设计任务 | AI 应用阶段 | | | AI 伦理模块（堂内讨论） | AI 技术模块（额外学时） | |
|---|---|---|---|---|---|---|---|---|
| 11–8（秋冬学期） | 64学时 | 约束性设计 | - | 设计 | - | 伦理 1：数据隐私保护问题、版权问题 | 技术 1：AI 通识与平台操作（人工智能基础知识，AIGC 工具基本操作） | 2学时 |
| | | | | | | | 技术 2：AIGC 提示词工程与图像生成（提示词逻辑、文生图、图生图、ControlNet 等工具） | 4学时 |
| 9–16（秋冬学期） | 64学时 | 系统性设计 | - | 设计 | - | 伦理 2：人机协同的权责边界 | 技术 3：LoRA 模型训练（图像数据预处理、模型训练与调优） | 4学时 |
| | | | | | | | 技术 4：ComfyUI 流程搭建（模块调用、样式控制、图纸生成流程构建） | 6学时 |
| 1–4（春夏学期） | 32学时 | 开放性设计 | - | 设计 | 表现 | 伦理 3：风格化与偏见问题 | | |
| 5–16（春夏学期） | 96学时 | 探究性设计 | 调研 | 设计 | 表现 | 伦理 4：模型偏误问题 | 技术 5：多模态工具整合（SD/Flux、LLM、机器/深度学习模型等的综合运用） | 4学时 |

## 1.4 学生反馈问卷设计

本研究采用三阶段结构化问卷（附录 1）系统评估 AI 融合建筑设计课程的教学效果。问卷设计涵盖三个核心维度：首先采集受试者的年龄、年级及 AI 课程基础等人口统计学特征，确保样本数据的代表性与有效性；其次运用李克特五级量表（1=完全无效，5=非常有效）定量分析 AI 工具在信息整理、逻辑分析、创意思维与设计表达四个维度的教学支持效能；最后通过半结构化访谈记录学生在调研、设计、表现三个阶段中 AI 工具的具体应用场景，并要求其对不同工具的阶段性效用进行优先级排序，同时设置 AI 技术持续使用意愿和伦理教学需求两个开放性议题。这种混合研究方法通过量化数据与质性资料的相互印证，为课程体系的持续优化提供了实证依据。

## 2. 教学过程及其作业成果

基于前文所述的 AI 教学模块分阶段融入模式，四个设计任务中 AI 应用呈现渐进特征：始于设计单阶段，延伸至设计-表现双阶段，终达全流程覆盖。为保持论述连贯性，成果展示根据 AI 技术融入的先后顺序，即按"设计-表现-调研"的顺序呈现（表 2）。

表 2 AI 技术应用与教学阶段匹配情况

| 任务阶段 | 阶段内容 | AI 应用重点 | AI 教学目标 |
|---|---|---|---|
| 前期调研 | 获取场地信息与背景资料 | 图像语义分割、图像特征提取、自然语言分析等 | 提升调研效率，培养数据意识 |
| 具体设计 | 生成与深化设计概念 | LLM 模型对话、AIGC 文生图/ | 激发创意思维，掌握迭 |

| | | 图生图、LoRA 模型训练等 | 代方法 |
|---|---|---|---|
| **后期表现** | 图纸渲染与表达优化 | AIGC 文生图/图生图、LoRA 模型训练、ComfyUI 工作流搭建等 | 提升表现效率，建立表达范式 |

## 2.1 设计阶段

具体设计阶段的 AI 技术教学以 AIGC 相关技术的基础与进阶应用为核心内容，通过系统性训练激发学生的创意思维并培养其设计迭代能力。考虑到学生在该阶段首次接触 AI 技术，课程采用循序渐进的技术教学模式。

在第一个设计任务中，教学重点在于 AI 辅助设计构思与草案优化。学生一方面通过 LLM 进行设计概念的拓展与优化，借助自然语言交互梳理设计原则并激发创新思路；另一方面利用 Stable Diffusion 的文生图和图生图功能优化自己的设计草案。在此过程中，学生通过提示词工程、参数调整等技术手段迭代生成的文本或图像，并基于专业判断对生成结果进行筛选与深化，从而显著提升方案初期的生成效率。随着教学深入，自第二个设计任务中期起，学生已可通过自定义 LoRA 模型的训练，逐步建立符合建筑学专业要求的标准化模型训练体系与模型库（图 2）。在后续的第三、第四个设计任务中，学生已具备自主搭建 ComfyUI 工作流的能力，将 AIGC 技术整合为定制化的设计流程，实现设计方法论的体系化沉淀与团队协作优化（图 3）。

需特别强调的是，涉及本阶段的 AI 伦理教学着重探讨人机协同的权责边界问题，应让学生意识到，AI 技术是协同、迭代，而非速成、替代。在教学实践中，我们引导学生建立三个关键认知维度：其一，在工具属性层面，强调 AI 作为设计思维的协同者而非主导者，需对生成内容保持审慎评估，不可盲目采纳；其二，在数据影响层面，揭示训练集偏差可能引发的设计趋同现象与文化失真风险，要求持续运用专业判断进行干预调整；其三，在应用伦理层面，明确 AI 辅助的合理作用范围，确保技术赋能聚焦于设计质量突破而非流程简化。这种教学模式不仅使学生获得 AI 协同设计的技术能力，更重要的是培育了其在智能化设计语境中的主体意识和辩证思维，最终实现人机优势互补的创新范式，使技术应用真正成为设计创新的助推器而非替代品。

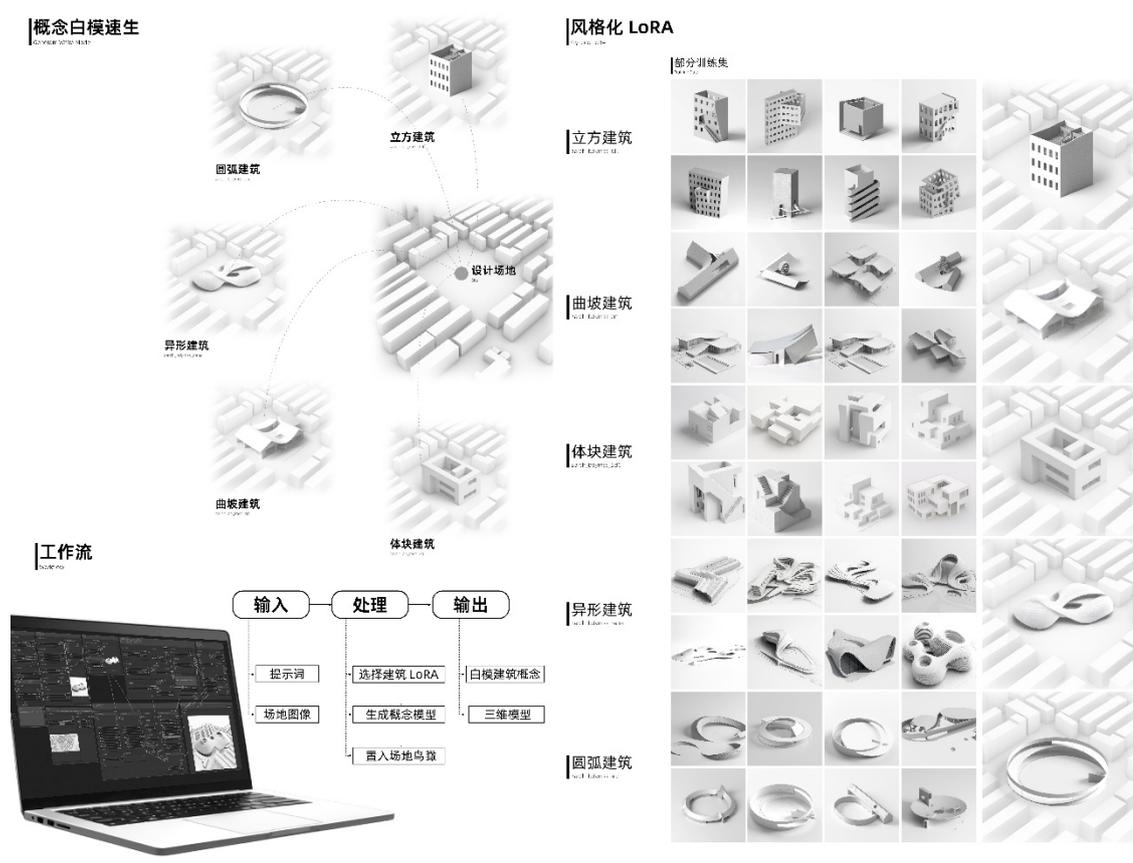

图 2 学生自研 LoRA 模型及 ComfyUI 工作流 — 在场地中生成多种概念模型可能性

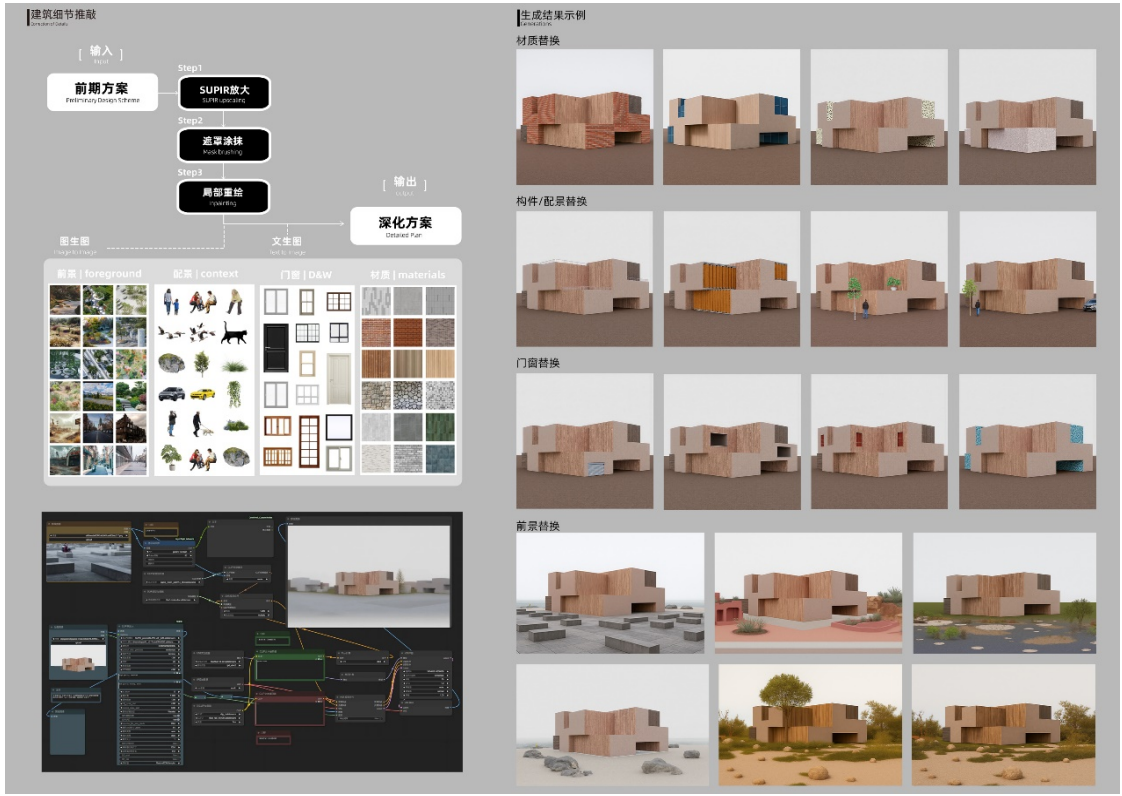

图 3 学生自研 ComfyUI 工作流 – 根据参考图快速测试不同细部效果

### 2.2 表现阶段

在后期表现阶段，学生运用 AIGC 技术优化最终设计方案的视觉呈现，显著提升了效果图的表达效率与质量。学生在此阶段将面临三大技术挑战：其一，建筑本体细节在输出图像中需保持一致性；其二，建筑空间及场地环境的语义关系需保证准确性；其三，建筑细部、材料质感及环境要素的比例需具备合理性。需要指出的是，以上难题在现有 AIGC 技术框架下尚未形成系统化的解决方案。因此，本课程将 AIGC 表现模块设置为高阶教学内容，安排在最后两个设计任务单元实施。为有效解决上述技术挑战，在流程控制层面，学生通过 ComfyUI 工作流集成 ControlNet、局部重绘、遮罩绘图、高清修复等专业化节点，实现对生成图像的精确调控（图 4）；在模型选择层面，学生需根据设计需求选取特征适配的 SDXL、Flux 或 DALL-E 3 等基础模型，提升输出图像的品质。

该阶段的 AI 伦理关注风格性与偏见问题，在风格自主性方面，通过对比分析不同基础模型的生成结果，使学生明确意识到算法可能带来的风格同质化风险，警惕对预设美学范式的无意识依赖；在文化偏见方面，重点剖析训练数据集局限性所导致的材质失真、地域特征误读等问题，培养学生通过专业素养进行辨别与修正的能力。需要特别强调的是，在人机协同的创作过程中，设计师必须始终保持主导地位——参数设置应服务于设计意图的表达，而非反向制约创作思维。

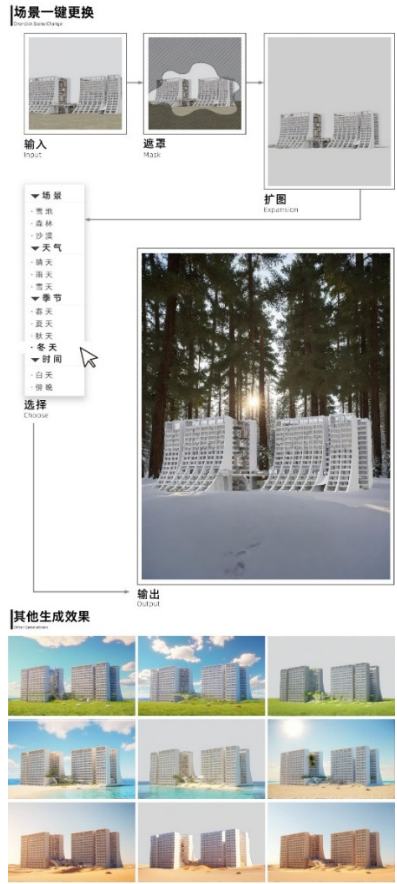

图 4 学生自研 ComfyUI 工作流 – 优化效果图场景

### 2.3 调研阶段

调研阶段虽在设计过程初期，但其涉及到的技术工具繁多，因而相关 AI 技术融入训练被安排在最后一个设计任务中。在建筑设计课教学中，任务调研与策划作为前期重要环节，其质量直接影响设计方案的合理性与创新性。然而，传统调研方法存在效率瓶颈，学生在有限时间内往往难以完成全面的数据采集与深度分析，导致前期调研多停留于定性层面。本次教学实验在前期调研阶段的 AI 技术教学目标是提升调研效率，培养数据意识。本次教学实验发现，多模态 AI 工具的介入显著提升了学生对于设计任务的调研效能，主要体现在以下三个方面：其一，借助大语言模型实现文献资料的智能检索与文本分析，辅助学生优化设计策划案的文本结构，提升前期设计建议的信息密度与逻辑性；其二，与二年级既有的参数化设计教学成果相衔接，同时整合 AIGC 技术，探索建筑方案的参数化生成与多方案可视化对比（图 6）；其三，将科研转化的预训练深度学习模型，如基于深度学习的图像语义分割模型（图 7）、特征提取模型（图 8）等，应用于教学实践，使学生能够快速获取场地地理环境的关键参数并作出定量结论。

在该阶段的随堂讨论中，模型偏误（Model Bias）问题作为 AI 伦理教学的核心议题被重点讨论。由于训练数据的局限性或算法设计缺陷，AI 模型可能产生系统性偏差，例如在语义分割任务中忽视图像数据本身的有偏性，或在文本

分析中过度强化模型训练数据库中数据权重偏高的设计范式。本课程通过让学生在实际操作中对比 AI 输出与人工调研结果的差异，使其直观认识到技术工具的局限性，并培养批判性使用 AI 的能力。这种技术融合的教学方法不仅显著提高了调研效率，更培养了学生的数据驱动设计思维，同时建立起对智能技术应用的辩证认知。

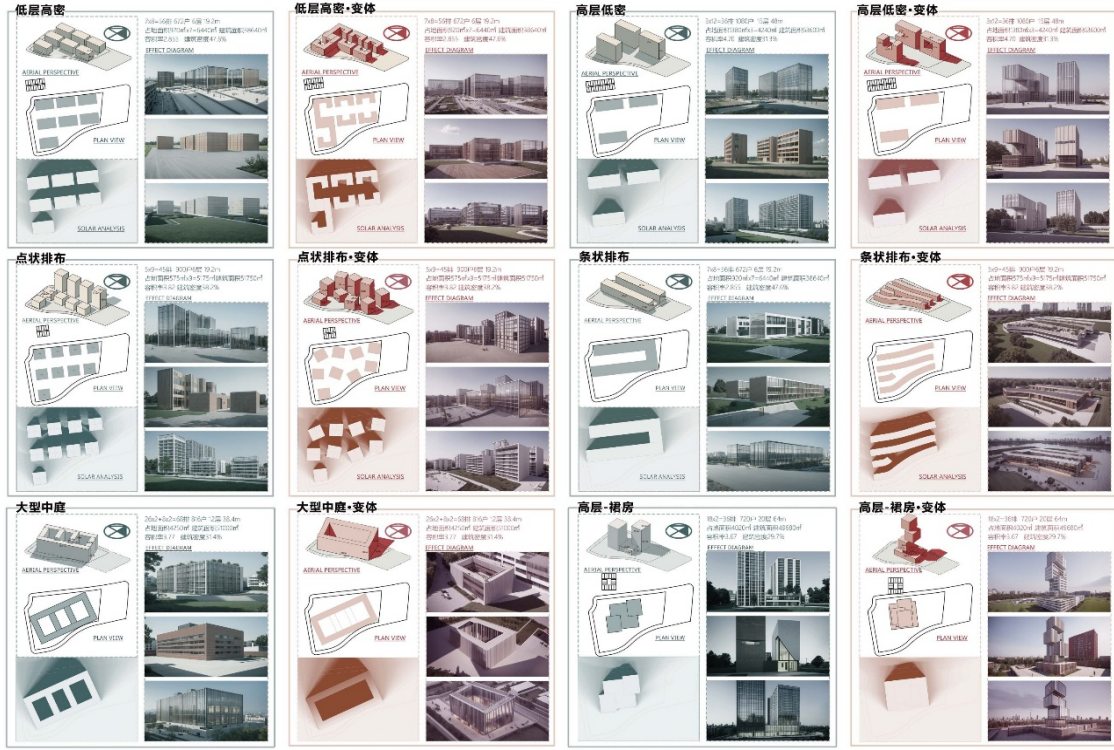

图 5 学生调研作业-应用参数化设计与 AI 技术进行方案"强排"并进行效果可视化

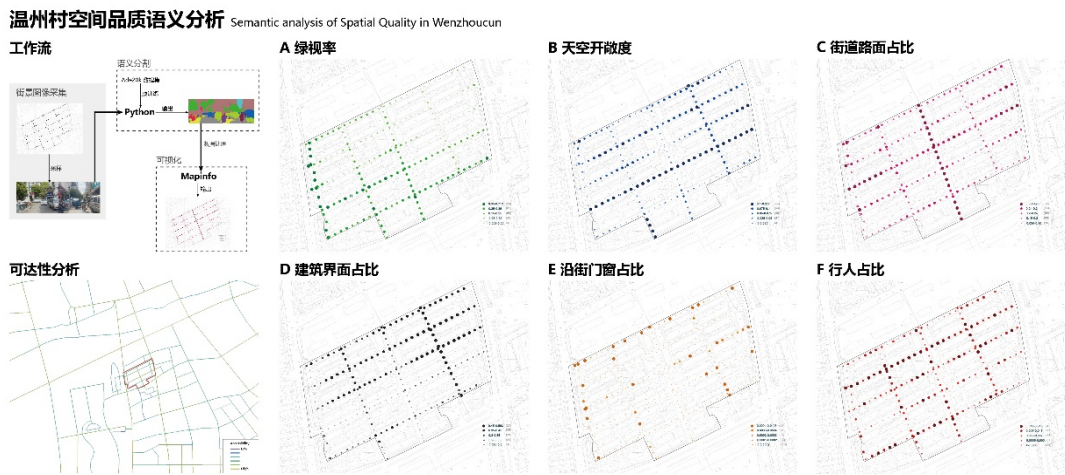

图 6 学生调研作业-应用深度学习模型（DCNN）开展场地街景图像语义分割

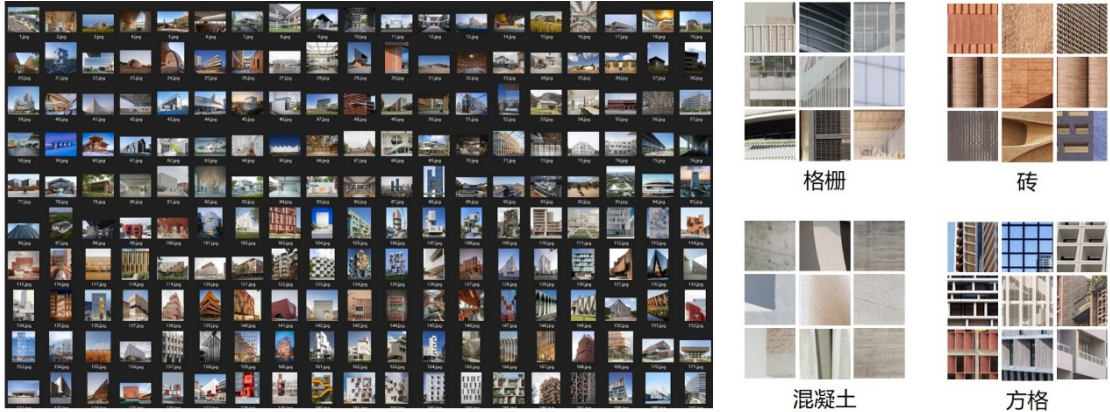

为建筑设计师提供建筑元素灵感库，快速、便捷地查阅建筑细部的各种做法。

图 7 学生调研作业-应用深度学习模型（Transformer）对谷德设计网热门建筑照片中的建筑细部进行智能聚类分析

### 2.4 问卷调研结果

本研究共回收有效问卷 70 份，样本性别分布均衡（男性 54.3%，女性 45.7%）。调查结果显示，学生的参与程度呈现梯度分布：24.3%仅完成听课，45.7%参与听课和跟练，30.0%完成实际设计应用。在课程收获方面，95.7%的受访者表示有所获益，其中 35.7%评价为"收获较多"（4 分），24.3%认为"充分受益"（5 分）。在课程中对 AI 工具的实际使用频率方面，13.7%的学生从未使用过，51.0%仅在课堂尝试过，29.4%多次用于过程讨论，5.9%在作业或图纸绘制中经常使用。（表 3）

在 AI 工具效能评价中（表 4），AI 工具的表现呈现显著差异：信息整理能力获得最高认可（84.1%评分≥4 分），逻辑分析能力次之（67.1%）；相比之下，AI 在创新思维（44.3%）和设计表达（44.3%）方面的支持效果相对有限，虽然平均评分仍高于中间值，但较大的标准差反映出学生在创造性应用方面存在明显认知差异。

AI 工具分阶段应用情况表明（表 5），AI 工具的使用呈现阶段性特征：具体设计阶段使用率最高（62.9%），前期调研阶段次之（55.7%），后期表现阶段最低（31.4%），表明其在前序设计流程中更具适配性。针对在不同阶段实际应用过相关工具的同学，问卷要求其对四类工具进行 AI 工具分阶段效能排序：大语言模型（LLMs）、多模态模型、AIGC 图像工具和自定义 AI 工作流（表 6）。在调研阶段，LLMs 在平均评分和首选率上均表现最优，表明学生普遍认可其在信息综合和早期概念探索中的价值；在设计阶段，各工具评分趋于均衡，虽然 LLMs 仍保持领先，但多模态模型和自定义 AI 工作流的评分逐渐提升，表明随着设计任务的推进，更复杂的工具变得愈发重要；在表现阶段，AIGC 图像工具虽获得最多首选，但其平均评分却偏低，反映出个体工具偏好分化的情况；值得注意的是，在此阶段自定义 AI 工作流的均值和排名已超越 LLMs，显示其在进阶专业任务中的潜力。

针对 AI 技术持续使用意愿和伦理教学需求两个开放性议题（表 7），问卷显示：在技术应用方面，82.8%的受访学生表现出积极态度，其中 25.7%计划在未来设计中扩大 AI 工具使用规模，57.1%持审慎乐观态度并期待技术进一步完善，

14.3%选择暂时观望，仅 2.8%因技术门槛或需求不匹配明确表示拒绝使用。在伦理教育形式上，65.7%的学生倾向采用非正式讨论或自主反思等柔性教学方式，显著高于支持传统课堂讲授的比例（28.6%），另有 5.7%认为无需专门设置伦理教学内容。

总体而言，问卷反馈揭示了 AI 工具在建筑设计任务中的两个核心特征：一是阶段依赖性，即 LLMs 在前期优势明显，而随着设计进程的推进，复杂 AI 工具的重要性逐步提升；二是工具敏感性，这既体现在不同工具在特定设计阶段呈现的差异化效能，也反映在学生个体对复杂 AI 工具使用的偏好分歧上，这种差异可能源于学生在技术掌握程度和应用需求方面的个体化差异。值得注意的是，问卷数据显示全体受访者均表现出对 AI 技术及其伦理议题的基本认知，而开放性问题的反馈结果进一步表明，学生不仅对当前 AI 技术的发展阶段保持着理性判断，同时也对伦理教育的形式和内容提出了个性化需求。这些发现可为建筑教育中 AI 技术的整合应用提供了重要参考。

表 3 问卷结果统计 – AI 工具学习与应用基本情况

| 问题 | 选项 | 数量 | 百分比 |
| --- | --- | --- | --- |
| 性别 | 男性 | 38 | 54.3% |
|  | 女性 | 32 | 45.7% |
| AI 课程参与度 | 未参加过 | 0 | 0.00% |
|  | 仅听课 | 17 | 24.3% |
|  | 听课并跟练 | 32 | 45.7% |
|  | 跟练并尝试在设计课中应用 | 21 | 30.0% |
| AI 课程获益程度 | 1：无用处 | 1 | 1.4% |
|  | 2：略有用处 | 2 | 2.9% |
|  | 3：收获尚可 | 25 | 35.7% |
|  | 4：收获较多 | 25 | 35.7% |
|  | 5：充分受益 | 17 | 24.3% |
| 在设计课程中使用 AI 工具的程度 | 1：完全未使用过 | 14 | 13.7% |
|  | 2：仅在课堂尝试过 | 52 | 51.0% |
|  | 3：多次使用并用于过程讨论 | 30 | 29.4% |
|  | 4：经常使用并用于作业/图纸中 | 6 | 5.9% |

表 4 问卷结果统计 – AI 工具效能评价

| 问题 | 平均值 (M) | 标准差 (SD) | 高分百分比 (4-5 分) |
| --- | --- | --- | --- |
| AI 工具具有强大的信息整理能力（例如对输入的大量文件资料的整理能力） | 4.28 | 0.84 | 84.1% |
| AI 工具具有强大的逻辑分析能力（例如基于输入的资料进行衍生分析的能力） | 3.86 | 0.94 | 67.1% |
| AI 工具具有强大的创新思维能力（能够生成突破常规的想法） | 3.46 | 1.02 | 44.3% |
| AI 工具具有强大的设计表达能力（包括图像表达和文字表现力） | 3.47 | 1.03 | 44.3% |

表 5 问卷结果统计 – AI 工具分阶段应用情况

| 设计阶段 | 数量 | 百分比 |
| --- | --- | --- |
| 前期调研 | 39 | 55.7 |
| 具体设计 | 44 | 62.9 |
| 后期表现 | 22 | 31.4 |

表 6 问卷结果统计 - AI 工具分阶段效能排序

| 工具类型 | 平均排名（调研） | 首选计数（调研） | 平均排名（设计） | 首选计数（设计） | 平均排名（表现） | 首选计数（表现） |
|---|---|---|---|---|---|---|
| 大语言模型（LLMs） | 1.35 | 25 | 2.07 | 15 | 2.47 | 4 |
| 多模态模型 | 1.74 | 14 | 2.16 | 14 | 2.64 | 3 |
| AIGC 图像工具 | 3.04 | 0 | 2.41 | 9 | 1.80 | 10 |
| 自定义 AI 工作流 | 3.59 | 0 | 2.82 | 6 | 2.14 | 5 |

表 7 问卷结果统计 - 未来应用预期与伦理教学态度

| 问题 | 选项 | 数量 | 百分比 |
|---|---|---|---|
| 未来应用预期 | 非常感兴趣，会使用并考虑以后会更多使用 | 18 | 25.7% |
| | 会使用，但认为目前 AIGC 工具不够成熟，会观望其发展 | 40 | 57.1% |
| | 暂时不考虑使用，但会持续关注 AIGC 工具的发展 | 10 | 14.3% |
| | 不会使用，考虑 AIGC 技术门槛过高 | 1 | 1.4% |
| | 不会使用，不看好/不认同 AIGC 工具 | 1 | 1.4% |
| 伦理教学态度 | 是的，有必要进行正式教学和课堂讨论 | 20 | 28.6% |
| | 有一定必要，但非正式讨论或自我反思即可 | 46 | 65.7% |
| | 没有必要 | 4 | 5.7% |
| | 不知道，对此没有概念 | 0 | 0 |

## 3. 总结与展望

### 3.1 教学改革的优势与实施经验总结

本次"AI+建筑学"教学实验在教案结构上的调整相对轻量，实质内容却实现了深度革新。所谓"轻量"，在于课程的整体框架、原有教学任务与师资结构基本维持不变，仅通过增设一位专任 AI 技术导师并引入模块化教学结构，实现 AI 教学内容的嵌入。这种结构安排在不干扰传统教学节奏的基础上，利用线上平台（如智云课堂与慕课课程）整合资源，将新增技术教学控制在全年不超过 20 学时之内。考虑到当前建筑设计课通常为小组指导制，师生比例约为 1:9，一个年级需同时配备多位指导教师。若要求所有教师均掌握最新 AI 工具并具备实际指导能力，在时间与培训资源有限的现实条件下存在较大挑战。因此，设置专任技术导师不仅有效缓解了师资技术掌握的差异性，也提升了 AI 模块教学的连续性与专业性，为高强度设计课教学提供了可持续的技术支持。对于本课程 5.0 学分的教学强度而言，这种控制课时增量、整合教学职责的方式在教学负担与教学效果之间达成了良好平衡，亦为其他高校提供了具备推广价值的参考样本。

所谓"革新"，在于课程设计与教学实施充分响应了 AI 技术的快速演进趋

势，展现出高度的前瞻性与适应能力。本次教学实验自 2023 年底构思，并于 2024 年 9 月正式启动，直至 2025 年 6 月完成一学年教学，期间恰逢多项 AI 核心工具密集发布：Stable Diffusion XL（2023 年 7 月）、DALL·E 3（2023 年 10 月）、Stable Diffusion 3（2024 年 2 月）、Flux（2024 年 8 月）、ComfyUI 重大版本更新（2024 年 12 月）、Grok 3（2025 年 2 月）、DeepSeek R1（2025 年 1 月）、OpenAI o3-mini（2025 年 1 月）、OpenAI o3 / o4-mini（2025 年 4 月）、OpenAI o3-pro（2025 年 6 月）等陆续上线。面对技术更新频率日益密集、工具系统持续扩展的现实情境，本课程通过设立专任 AI 技术导师、引入弹性模块化结构，建起一套可随技术动态调整的教学机制，在教学实验期间持续更新迭代讲座内容，确保学生始终处于智能设计工具的技术前沿。该机制不仅提升了课程的技术响应力，也显著增强了教学体系的生命力与可持续性，真正实现了"以变应变"的教学创新逻辑。

### 3.2 教学反思与优化方向

在教学实践过程中，仍存在若干值得反思与持续优化的问题。从技术层面来看，学生对 AI 工具的接受度不一，且部分操作流程涉及编程思维，对建筑学背景的学生构成挑战。部分学生虽能在设计初期借助 AI 完成概念草图与图像生成，但在面对空间组织、功能布局与结构逻辑等复杂环节时，AI 协同能力有限，难以支撑深层次设计推演。在课时有限的情况下，学生往往难以兼顾技术适配与设计深化，影响整体学习成效。这在很大程度上源于当前 AI 工具尚缺乏面向建筑复杂问题的高适配性。尽管存在技术瓶颈，本教学组主张本次教学改革的核心并非短期内实现 AI 深度协同，而在于培养学生对 AI 的敏感性与适应力，引导其在学习过程中主动接触、尝试并理解新兴工具，推动其认知结构从"传统主导"向"智能协同"转化。此过程虽非一蹴而就，却可为未来实现更高水平的人机共创奠定基础。

从思想层面而言，对 AIGC 所带来的"趋同化"风险，师生普遍表达担忧。其高度模式化与风格复用机制，存在令新一代的审美判断陷入自我循环、趋于停滞的隐患。然而，艺术史反复印证，每一次重大技术革新——无论是透视法的引入，还是摄影术的兴起——都深刻重塑了人类对"美"的认知方式[18]，AIGC 在当下也正以"风格规约—变式演化"的方式，推动未来视觉意象逻辑的重构[19]。正是这种技术与美学的辩证关系，使得对 AI 伦理的持续探讨显得尤为重要。《建筑设计Ⅲ/Ⅳ》课程虽未单独设置 AI 伦理模块课时，但已明确要求各指导教师在教学中主动引导学生讨论人机协同责任、模型偏误、训练数据偏见等关键议题。然而，在实际操作中，由于教师个人理解与立场存在差异，部分讨论可能易受主观倾向影响，难以保障学生形成独立、多元的判断能力。针对这一问题，未来可通过建设 AI 伦理教学参考资料库，引入学界相关研究成果作为统一支撑材料，并辅以教学组定期研讨机制，以提升教学一致性与引导策略，从而更有效地在"AI+建筑学"的语境中培养学生的批判意识与设计判断力。

### 3.3 "AI+建筑学"路径下的建筑学创新人才培养思考

"AI+建筑学"模式不仅是技术教学的拓展，更是高等教育中建筑学人才培养理念的延展与实践。浙江大学建筑学系强调"全面养成"，致力于培养具备

宽阔知识背景、多元适应能力和全球竞争力的创新复合型卓越人才。本课程以此为导向，将 AI 知识系统性嵌入核心设计课教学，构建"技术理解—工具使用—思辨强化"的能力结构，引导学生主动适应前沿技术，同时始终保持对设计主导权的掌控。

AI 生成过程本质上充满"选择"与"判断"，学生需在多个生成结果中做出筛选，这一过程要求其在技术支持下强化专业判断，而非被动接受 AI 输出。课程强调，设计师应逐步建立起将 AI 作为协同对象而非替代工具的认知框架，理解人机协同的边界、价值与责任。在高频交互训练中，学生既掌握了 AI 工具的操作，也锻炼了甄别、筛选与再创造的能力。通过技术与伦理双模块并行设计，课程推动学生在掌握 AI 应用的同时，建立起理性使用与深度反思并重的认知体系。

### 3.4 小结

综上，2024–2025 学年《建筑设计Ⅲ/Ⅳ》课程的"AI+建筑学"教学实验，在保留建筑设计教育核心价值的基础上，验证了人工智能技术与伦理双模块在建筑设计课程中的有效嵌入路径，并形成以下主要结论：首先，模块化嵌入的教学改革方式被证明具有良好的可行性。在不改变原有课程结构的前提下，将 20 学时 AI 技能训练与课堂伦理讨论相结合，能够在有限教学资源下同步提升学生的技术能力与批判性认知，为 AI 融合建筑教育提供了可复制的模式。其次，专职技术导师的配置是保障实施成效的关键。其持续介入不仅确保了 AI 教学内容的前沿性与系统性，也显著提升了学生的技术掌握深度与应用水平。此外，问卷调研显示学生对 AI 工具的使用具有阶段依赖性与工具敏感性。前期调研与具体设计阶段的积极性最高，对大型语言模型掌握较为稳定；而在图像生成工具与自定义工作流的应用上差异明显，反映出技术掌握与认知准备度的分层。同时，AI 伦理教学不仅促进了数字化适应能力、策略性设计思维与批判性判断力的提升，还引导学生在设计过程中反思作者身份、算法偏差与责任归属，将 AI 视为需引导与合作的"共创者"而非不可控的"黑箱工具"。最后，完善的技术教学组织方案与平台支撑是保障成效的关键。依托在线学习平台、云计算资源以及科研产出的预训练模型和定制化工作流，课程在教学实践中不仅产出了 LoRA 模型包、ComfyUI 工作流和 AI 数据集等创新成果，验证了"AI+建筑学"的协同增效机制，还初步建立了双轨并行的教学模式，呈现出清晰的实施路径与良好的扩展潜力，为"AI+"时代建筑教育的体系优化与跨院系推广提供了可复制的参考样本。

**参考文献**

# Bibliography

hadid-architects-builds-winner-proposals-with-ai-enterprise-network-qs7m7txwz.

[4] BAŞARIR L. Modelling AI in architectural education[J]. Gazi University Journal of Science, 2022, 35(4): 1260-1278.

[5] 庄惟敏, 孙一民, 汪孝安, 等. "坚守·创新——应对建筑学的时代挑战"笔谈[J]. 新建筑, 2025(1): 158-166.

[6] ABDULLAH H K, HASSANPOUR B. Digital design implications: A comparative study of architecture education curriculum and practices in leading architecture firms[J]. International Journal of Technology and Design Education, 2021, 31(2): 401-420.

[7] JIN S, TU H, LI J, 等. Enhancing architectural education through artificial intelligence: A case study of an AI-assisted architectural programming and design course[J]. Buildings, 2024, 14(6): 1613.

[8] 闵嘉剑, 于博柔, 张昕. 生成式人工智能时代的设计教学探索——以清华大学"ai 生成式影像"课程为例[J]. 建筑学报, 2023(10): 42-49.

[9] 东南大学建筑学院. 智能 AI 算法助力课程设计——可变住宅 | 东南大学建筑学专业四年级课程设计[EB/OL]. (2024)[2025-06-25]. https://arch.seu.edu.cn/2024/0604/c9122a492630/page.htm.

[10] 陈雷, 张伶伶, 陈子墨, 等. 基于图像生成式人工智能的"人—机—人"建筑设计教学模式研究[J/OL]. 建筑师, 2024[2025-08-15]. https://kns.cnki.net/KCMS/detail/detail.aspx?dbcode=CAPJ&dbname=CAPJLAST&filename=JZSS20240626003.

[11] GSD. Generative AI in teaching and learning at the GSD[EB/OL]. (2024)[2025-06-25]. https://www.gsd.harvard.edu/resources/ai/.

[12] GSD. AI, machine learning, and the built environment - harvard graduate school of design executive education[EB/OL]. (2025)[2025-06-25]. https://execed.gsd.harvard.edu/programs/artificial-intelligence-built-environment/?utm_source=chatgpt.com.

[13] GSAPP. Spatial AI[EB/OL]. (2025)[2025-06-25]. https://www.arch.columbia.edu/courses/14172-4740?utm_source=chatgpt.com.

[14] GSAPP. Ai for existing buildings[EB/OL]. (2022)[2025-06-25]. https://www.arch.columbia.edu/courses/13583-5539-ai-for-existing-buildings?utm_source=chatgpt.com.

[15] TU DELFT. MOOC: AI in architectural design: introduction | TU delft online[EB/OL]. (2025)[2025-06-25]. https://online-learning.tudelft.nl/courses/ai-in-architectural-design/?utm_source=chatgpt.com.

[16] 吴越, 许伟舜, 孟浩. 从链条到生态——浙江大学建筑学系的数字化课程体系改革[J]. 高等建筑教育, 2024, 33(1): 67-75.
16

## 图表来源

图 1，表 1-7： 作者整理、绘制

图 2-7：学生作业

## 文末备注

## 附录 1

### 1. 基本信息

1.1 您的姓名:

1.2 您的学号:

1.3 您的性别:

1.4 您是否参加过浙江大学建筑设计课程中的 AI 教学模块？

□未参加过　　　　□仅听课　　　　□听课并跟练　　　　□跟练并尝试在设计课中应用

1.5 请评价参加该课程对您设计学习的受益程度:

□1:无用处　　□2:略有用处　　□3: 收获尚可　　□4: 收获较多　　□5: 充分受益

### 2. AI 工具有效性评估（1=完全不同意，5=完全同意）

2.1：AI 工具具有强大的信息整理能力（例如对输入的大量文件资料的整理能力）。

2.2：AI 工具具有强大的逻辑分析能力（例如基于输入的资料进行衍生分析的能力）。

2.3：AI 工具具有强大的创新思维能力（能够生成突破常规的想法）。

2.4: AI 工具具有强大的设计表达能力（包括图像表达和文字表现力）。

### 3. AI 融合教学的描述性反馈

3.1 您在设计课程中尝试使用 AI 工具的程度如何？

□完全未使用过　　□仅在课堂尝试过　　□多次使用并用于过程讨论　　□经常使用并用于作业/图纸中

3.2 在建筑设计课程中，您在哪些阶段使用 AI 工具？（多选题）

□前期调研　　□具体设计　　□后期表现

3.3 分别对不同设计阶段使用的 AIGC 工具实用程度排序(1-4)：

（只有在 3.2 题中选择了对应阶段的受访者，才需要完成此排序）

| 阶段 | AI 工具 | | | |
| --- | --- | --- | --- | --- |
| 3.3.1 前期调研 | □大语言模型 | □多模态模型 | □AIGC 图像工具 | □自定义 AI 工作流 |
| 3.3.2 具体设计 | □大语言模型 | □多模态模型 | □AIGC 图像工具 | □自定义 AI 工作流 |
| 3.3.3 后期表现 | □大语言模型 | □多模态模型 | □AIGC 图像工具 | □自定义 AI 工作流 |

3.4 您在后续设计中是否还会/考虑使用 AIGC 工具？

□非常感兴趣，会使用并考虑以后会更多使用　　□会使用，但认为目前 AIGC 工具不够成熟，会观望其发展　　□暂时不考虑使用，但会持续关注 AIGC 工具的发展　　□不会使用，考虑 AIGC 技术门槛过高　　□不会使用，不看好/不认同 AIGC 工具

3.5 你认为是否有必要对 ai 伦理进行正式教学并展开课堂讨论？

□是的，有必要进行正式教学和课堂讨论　　□有一定必要，但非正式讨论或自我反思即可　　□没有必要　　□不知道，对此没有概念